\documentclass[twocolumn,showpacs,amsmath,amssymb,pra]{revtex4}

\usepackage{graphicx}
\usepackage{dcolumn}
\usepackage{bm}
\usepackage{units}

\usepackage[plainpages=false, bookmarks=false, pdftoolbar=true, %
	pdfmenubar=true, pdffitwindow, %
	pdftitle={Radio-frequency driven dipole-dipole interactions in spatially%
	separated volumes}, pdfauthor={Atreju Tauschinsky}, %
	colorlinks=true, pdfhighlight=/O, pdfstartview=FitH, linkcolor=blue, %
	citecolor=red, urlcolor=blue]{hyperref}

\usepackage{amsmath}

\begin{document}

\title{Radio-frequency driven dipole-dipole interactions %
	in spatially separated volumes}

\author{Atreju Tauschinsky}%
\author{C. S. E. van Ditzhuijzen}%
\author{L. D. Noordam}%
\author{H. B. van Linden van den Heuvell}%
\email{H.B.vanLindenvandenHeuvell@uva.nl}
\affiliation{Van der Waals-Zeeman Institute, University of Amsterdam, %
Valckenierstraat 65, 1018 XE Amsterdam, The Netherlands.}

\date{\today}	

\begin{abstract}
Radio-frequency (rf) fields in the MHz range are used to induce resonant energy 
transfer between cold Rydberg atoms in spatially separated volumes. After laser 
preparation of the Rydberg atoms, dipole-dipole coupling excites the 
49s atoms in one cylinder to the 49p state while the 41d atoms in the second 
cylinder are transferred down to the 42p state. The energy exchanged between the 
atoms in this process is \unit[33]{GHz}. An external rf-field brings 
this energy transfer into resonance. The strength of the interaction has been 
investigated as a function of amplitude (\unitfrac[0--1]{V}{cm}) and 
frequency (\unit[1--30]{MHz}) of the rf-field and as a 
function of a static field offset. Multi-photon transitions up to fifth order 
as well as selection rules prohibiting the process at certain fields have been 
observed. The width of the resonances has been reduced compared to earlier 
results by switching off external magnetic fields of the magneto-optical trap, 
making sub-MHz spectroscopy possible. 
All features are well reproduced by theoretical calculations taking the 
strong ac-Stark shift due to the rf-field into account.
\end{abstract}


\maketitle

\section{Introduction}
Rydberg atoms have appealing properties for the study of atom-atom 
interactions; in particular their lifetime is very long and their large 
(transition) dipole moments make them extremely sensitive to electric fields. 
Rydberg atoms can be used as sensitive probes of the environment or vice versa: 
the sensitivity to electric fields makes it easy to control their behaviour 
through the application of external fields.

Resonant interactions between dipoles are of great interest in a broad range of 
fields and manipulating the coupling between dipoles is very useful in such 
diverse areas of physics as nanophotonics and quantum computation. 
Dipole-interacting Rydberg atoms are under active study by various groups as 
reported in Ref. 
\cite{Afrousheh2004, Zhang1994, Li2005, Vogt2006, Westermann2006, Johnson2008}, 
as an ideal system to investigate such interactions.

Previously \cite{Ditzhuijzen2008} we investigated the dipole-coupling between 
Rydberg atoms in two separated volumes as a function of static electric field, 
minimum particle distance and interaction time. In the present paper we report 
how a radio-frequency field can provide additional control over the interaction 
strength. Radio-frequency and microwave fields have previously been used to 
investigate the properties of both isolated and interacting Rydberg atoms e.g.\ 
in \cite{Bohlouli-Zanjani2007, Gallagher1982, Zhang1994, %
Bloomfield1986, Pillet1987}. The techniques presented in this article will 
allow precision spectroscopy on 2-atom states and might be useful in quantum 
computation schemes to minimize electric field inhomogeneities which influence 
the dephasing of qubits implemented as Rydberg atoms 
\cite{Bohlouli-Zanjani2007}.

\begin{figure}[htb]
	\centering
	\includegraphics[width=\columnwidth]{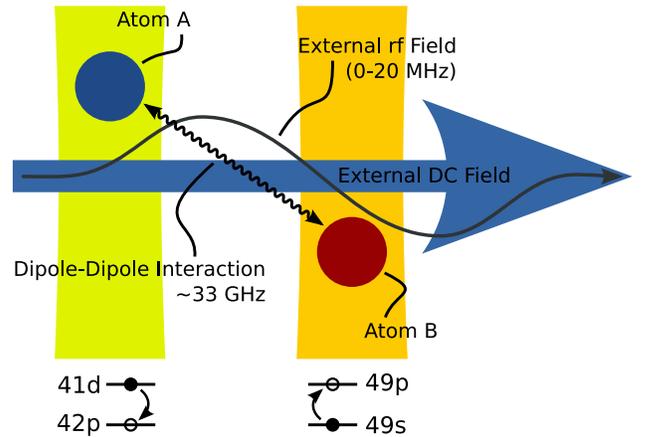}
	\caption{(Color online) Schematic of the experiment. Fields are represented 
	by the labelled arrows. The two upright areas symbolize the laser beams 
	creating the Rydberg atoms. Only two atoms are shown for simplicity while 
	in reality there are some tens of atoms in total. The Rydberg volumes are 
	typically \unit[25]{$\mu$m} apart at a width of \unit[15]{$\mu$m.} The 
	change of states of the Rydberg atoms is also depicted.}
	\label{fig:comic}
\end{figure}

In the experiment presented in this paper we create two spatially separated 
volumes of Rydberg atoms by exciting cold ground-state atoms from a 
magneto-optical trap. This is achieved by focusing two dye lasers tuned to 
different Rydberg states into the trap (see Fig. \ref{fig:comic}). This 
procedure creates a few tens of Rydberg atoms in each laser focus. After 
letting the atoms interact for a fixed period of time, they are field-ionized, 
providing information on the states present. From this one can determine the 
number of atoms in the initial and final states, and hence the fraction of 
atoms having interacted. The interaction is influenced by external fields 
during the interaction time.

Fig. \ref{fig:comic} shows a schematic picture of the experiment. The photon 
associated with the dipole-dipole interaction is symbolized by the arrow 
between the two atoms. The other arrows represent external rf- and static 
electric fields. During the interaction time the interaction is tuned into 
resonance by these externally applied fields. An additional external field due 
to black body radiation is always present in the experiment.

\begin{figure}[htb]
	\centering
	\includegraphics[width=\columnwidth]{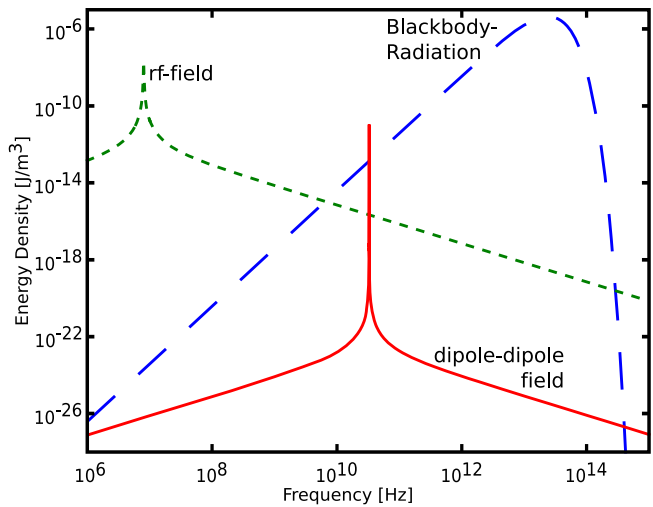}
	\caption{(Color online) Energy density spectrum of the three different 
	radiation fields, to which the atoms are exposed. The blue line 
	(long dashes) represents the always and everywhere present black body 
	radiation, depicted for \unit[300]{K}. Resonances of the vacuum chamber and 
	the metallic field plates are not taken into account for this illustration.
	The red line (solid) represents the photon that goes from one atom to the 
	other, due to the dipole-dipole interaction at \unit[25]{$\mu$m}. The line 
	shape is lorentzian based on experimental parameters. The green line 
	(dashed) represents the oscillating field that is externally applied.}
	\label{fig:energydensity}
\end{figure}

Fig. \ref{fig:energydensity} shows an illustration depicting the order of 
magnitude of the typical energy density of all time-dependent fields. Note that 
the energy density has been multiplied by 
$\omega$ giving units of \unitfrac{J}{m$^3$} as this leads to constant 
intervals d$\omega$ in the logarithmic plot. Also note that the scale of the 
y-axis covers more than 20 orders of magnitude. 
This shows that while there is a contribution of the black body-radiation in the 
background, the contribution of the dipole-dipole and rf-fields is much larger 
in the relevant frequency ranges. 

The process in our experiment can be interpreted as amplitude modulation of the 
dipole-field by the external rf-field. Surprisingly, the amplitude of the 
carrier wave (dipole-dipole radiation) is much smaller than the amplitude of 
the modulating rf-field. 

\section{Radiofrequency Assisted Dipole-Dipole Interactions}
The initial states chosen in this experiment are 41d and 49s, created in 
separate volumes. They exchange a photon in a resonant dipole-dipole 
interaction forming the states 42p and 49p in one of the processes
\begin{subequations}
\begin{align}
	41\mathrm{d}_{\frac{3}{2},\frac{1}{2}} + 49\mathrm{s}_{\frac{1}{2},%
	\frac{1}{2}} &\rightarrow 42\mathrm{p}_{\frac{1}{2},\frac{1}{2}} + %
	49\mathrm{p}_{\frac{3}{2},\frac{1}{2}}\\
	41\mathrm{d}_{\frac{3}{2},\frac{1}{2}} + 49\mathrm{s}_{\frac{1}{2},%
	\frac{1}{2}} &\rightarrow 42\mathrm{p}_{\frac{1}{2},\frac{1}{2}} + %
	49\mathrm{p}_{\frac{3}{2},\frac{3}{2}}
\end{align}
\label{eq:reaction2}
\end{subequations}
where the notation denotes n$\ell_{j, |m_j|}$. Both the 41d$_{\frac{3}{2}}$ and 
the 49p$_{\frac{3}{2}}$ states split in an electric field into two substates 
$|m_j| = 1/2$ and $|m_j| = 3/2$. Note that the initial 
41d$_{\frac{3}{2},\frac{3}{2}}$-state is not excited by the laser, because of 
its polarization, so it does not play a role in our experiment and was ignored 
in Eq. \eqref{eq:reaction2}. The approach of using four different states 
makes it possible to control the distance between the atoms, by having separate 
volumes for the two different initial states. A position-resolved detection 
technique is then not necessary; state selective detection of the final 
49p-state only will probe the dipole-dipole interaction. 

\begin{figure}[ht]
	\centering
	\includegraphics[width=\columnwidth]{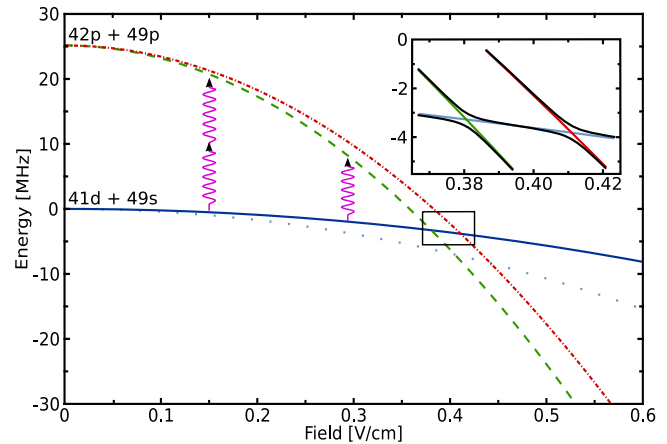}
	\caption{(Color online) Schematic representation of two-atom energy versus 
	electric field. An external rf-field of \unit[10]{MHz} is symbolized by the 
	arrows. At a purely static field of \unitfrac[0.38]{V}{cm} and 
	\unitfrac[0.41]{V}{cm} the dipole-dipole transition is resonant. At 
	\unitfrac[0.3]{V}{cm} it is brought into resonance by one rf-photon of 
	\unit[10]{MHz}, while at \unitfrac[0.15]{V}{cm} two such photons bring the 
	system into resonance. The inset shows the avoided crossings due to 
	dipole-dipole interaction at a separation of \unit[25]{$\mu$m}.}
	\label{fig:energy}
\end{figure}

The binding energies and polarizabilities of the states from Eqs. 
\eqref{eq:reaction2} are obtained from a calculated Stark map, using the 
Numerov method \cite{Zimmerman1979} and making use of the Ru\-bi\-di\-um-85 
quantum defects given in \cite{Li2003,Han2006,Afrousheh2006}. In Fig. 
\ref{fig:energy} the sum of the binding energies of the two atoms is plotted 
versus the electric field. Both the initial state (41d + 49s) as well as the 
final state (42p + 49p) undergo a purely quadratic Stark shift, which is much 
stronger for the final pp-state than for the initial sd-state. This is caused 
by the strong Stark shift of the 49p-state. 

For the dipole-dipole transfer to become resonant the energy difference between 
the initial and final two-atom state has to be zero. This energy difference can 
be written as 
\begin{equation}
	W=W_0-\frac{1}{2}\alpha F^2
	\label{eq:energy}
\end{equation}
where $W_0$ is the zero-field energy difference, \unit[$W_0 = 25.15(13)$]{MHz}, 
$F$ the total electric field and $\alpha$ the difference polarizability, i.e.\ 
the difference in polarizability between the initial and the final two-atom 
state. The nonlinear term in the field is due to the purely quadratic Stark 
shift of the energy levels. The energy difference at zero field ($W_0$) is 
already very small, which is one of the reasons we have chosen this particular 
process. The difference polarizabilities are 
$\alpha_\mathrm{a}$~=~\unitfrac[347.04(4)]{MHz}{(V/cm)$^2$} for process 
(\ref{eq:reaction2}a) and 
$\alpha_\mathrm{b}$~=~\unitfrac[297.40(4)]{MHz}{(V/cm)$^2$} for process 
(\ref{eq:reaction2}b); this makes the process resonant at the static fields 
\unitfrac[$F_\mathrm{a} = 0.3807(15)$]{V}{cm} and 
\unitfrac[$F_\mathrm{b} = 0.4113(16)$]{V}{cm} respectively.

For non-zero electric fields the quantum number $\ell$ does not describe a 
proper eigenstate of the system. We label those states as having the angular 
momentum quantum number $\ell$ that connect adiabatically to the zero-field 
states with that quantum number. However at an electric field of 
\unitfrac[0.4]{V}{cm} the mixing in of other $\ell$-states is still negligible. 
 
The strength of the dipole-dipole interaction, and therefore half the size of 
the avoi\-ded crossing is given in atomic units by
\begin{equation}
	V = \frac{\bm{\mu_1 \cdot \mu_2} - %
	3 (\bm{\mu_1 \cdot \hat R})(\bm{\mu_2 \cdot \hat R}) }{R^3}
	\label{eq:pot2}
\end{equation}
where $\bm{R}$ is the distance vector between the interacting particles, 
$\bm{\mu_1}$ is the dipole moment of the $41\mathrm{d} \rightarrow 
42\mathrm{p}$ transition and $\bm{\mu_2}$ the dipole moment of the 
$49\mathrm{s} \rightarrow 49\mathrm{p}$ transition. The dipole moments are 
rather large, of the order of \unit[1000]{a$_0$e}, which is another advantage 
of this particular process eqn.~\eqref{eq:reaction2}. The exact value and 
polarization of each dipole $\mu$ depends on the $\Delta m_j$ of the 
transition. The dipole is either linearly polarized ($\bm{\mu} = \mu_z$) or 
circularly polarized $(\bm{\mu} = \mu_x \pm i\mu_y)$. For the field strength of 
the oscillating dipoles we typically obtain \unitfrac[34]{$\mu$V}{cm} at a 
separation of \unit[25]{$\mu$m}. We can consider the dipole-dipole interaction 
process as the exchange of a photon from one atom to the other. This photon has 
an energy of 32.8 GHz. The inset in Fig. \ref{fig:energy} shows the avoided 
crossing due to the dipole-dipole interaction for this separation. The 
frequency of the quantum beat oscillation of the system (or the Rabi 
oscillation of the single atoms) is 
\unit[$\omega_{QB} = 2V \approx 2\pi\times200$]{kHz}.

In the presence of an external rf-field the sd-state can couple to the pp-state 
even far away from the static field resonance, as symbolized by the arrows in 
Fig. \ref{fig:energy}. This is the case if the energy of the rf-photon equals 
the energy difference between the sd- and pp-state. The energy difference of 
those states for an rf-field is
\begin{equation}
W = W_0 - \frac{1}{2}\alpha\left(F_S + F_{\mathrm{rf}}\sin{\omega t}\right)^2
\label{eq:timedepenergy}
\end{equation}
where $F_S$ is the static field offset and $F_{\mathrm{rf}}$ the amplitude of 
the radiofrequency field. As long as $\omega_{QB} \ll \omega$ we can average 
Eq. \ref{eq:timedepenergy} over time \cite{Zhang1994}; this yields
\begin{equation}
	\label{eqn:energydiff}
	W = W_0 - \frac{1}{2}\alpha\left(F_S^2 + \frac{1}{2}F_{\mathrm{rf}}^2\right)
\end{equation}
and we define an effective field
\begin{equation}
	F_{\mathrm{eff}}=\sqrt{F_S^2+\frac{1}{2}F_{\mathrm{rf}}^2}
\end{equation}
This effective field is defined such that the Stark shift for a static field 
equal to $F_{\mathrm{eff}}$ is the same as the time-averaged Stark shift of the 
time-dependent field. The rf-field brings the dipole-dipole interaction into 
resonance, if an integer number of rf-photons matches the energy difference W 
between the states, so
\begin{equation}
	N \omega = W_0 - \frac{1}{2}\alpha F_{\mathrm{eff}}^2
	\label{eqn:res}
\end{equation}
where $N$ is the number of photons involved in the transition and $\omega$ is 
the frequency of the external rf-photon. Note that in the case of a purely 
linear Stark shift the behavior is essentially different. In that case the 
time-average of the rf-field becomes zero and only the static field $F_S$ is 
relevant. The term $\frac{1}{4}\alpha F_{rf}^2$ in Eq. \eqref{eqn:energydiff} is 
also called the ac-Stark shift.

\section{Improved Experimental Setup}
In this experiment a similar setup to measure dipole-dipole transitions  is 
used as in earlier experiments described in 
\cite{Ditzhuijzen2008, Ditzhuijzen2006}. A Magneto-Optical Trap (MOT) is used 
to create a cold cloud of $^{85}$Rb atoms. Rubidium atoms from a dispenser 3 cm 
away are trapped and cooled by three orthogonal pairs of counter-propagating 
laser beams in a quadrupole magnetic field. The cold atom cloud contains 
typically several \unit[$10^7$]{atoms} at a temperature well below 
\unit[300]{$\mu$K}. Charged particles emitted by the dispenser are deflected by 
statically charged plates in front of the dispenser. The background pressure is 
below \unit[$3\times10^{-8}$]{mbar}. These conditions ensure that the atoms in 
the MOT do not move on the relevant timescales of the experiment. 

Rydberg atoms are created by \unit[8]{ns} laser pulses at \unit[594]{nm} in a 
two-photon process from the cold ground-state Rb atoms. There are two separate 
sets of Nd:YAG-pumped dye lasers, one which excites atoms to the 49s state and 
one which excites to the 41d state. The laser creating the 49s atoms has a 
linewidth of \unit[0.017(11)]{cm$^{-1}$} and a pulse energy of 
\unit[3.0(1)]{$\mu$J}; the laser creating 41d atoms has a linewidth of 
\unit[0.21(2)]{cm$^{-1}$} and a pulse energy of \unit[5.0(1)]{$\mu$J}. 
The repetition rate of both lasers is \unit[10]{Hz}.
Each can be focused into a different region of the MOT to create two spatially 
separated, cigar-shaped volumes of Rydberg atoms within the magneto-optical 
trap. The focus of the laser creating the 49s atoms can be moved laterally 
within the trap by use of a stepper motor. The beam separation can accurately 
be measured in a two-photon mixing process using one photon from each beam. For 
this measurement we detune the laser creating the 49s atoms by \unit[20]{GHz} 
to the blue and make sure that the laser pulses overlap in time. This leads to 
the excitation of the 44d state as described in \cite{Ditzhuijzen2008}. As 
reported there the diameter of the Rydberg volumes is \unit[11.6(0.4)]{$\mu$m} 
for 49s- and \unit[16.3(0.5)]{$\mu$m} for 41d-atoms.

\begin{figure}[ht]
	\centering
	\includegraphics[width=\columnwidth]{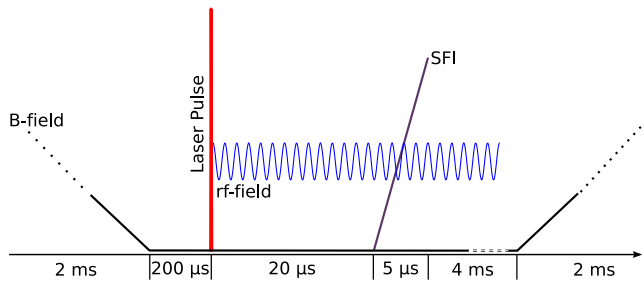}
	\caption{(Color online) Schematic of the timing of one experimental cycle. 
	First the magnetic field is switched off, a laser pulse excites Rydberg 
	states which then are field ionized after \unit[20]{$\mu s$} interaction 
	time. Finally the magnetic field is switched on again.}
	\label{fig:timing}
\end{figure}

The MOT is situated between two circular stainless steel plates with a diameter 
of \unit[5.5]{cm} which are spaced \unit[2.50(5)]{cm} apart. Both plates are 
perforated by a \unit[14]{mm} hole in the center to allow ionization products 
to pass through, as well as to enable optical access to the MOT. These plates 
serve to generate the static- and radio-frequency fields used to manipulate the 
dipole-dipole interaction as well as for the field ionization. 

The manipulating fields are generated by an Agilent 33250A Arbitrary Waveform 
Generator attached to one of the plates. The amplitude of these external fields 
is in the range of \unitfrac[0--1]{V}{cm} and the frequency of the external 
rf-field is in the range of \unit[1--30]{MHz}. The polarization of the rf-field 
is always parallel to the separation between the Rydberg volumes as they 
are separated along the symmetry axis of the plates.

The other plate is connected to a home-built fast high-voltage pulse generator 
to provide a negative field ramp that ionizes the Rydberg atoms after the 
interaction time. The ramp rises from 0 to 150 \unitfrac{V}{cm} in 5 $\mu$s. 
Electrons are detected on a Hamamatsu Multichannel Plate (MCP). Since each 
state ionizes at a different electric field the signal arises at a different 
time for different states \cite{Gallagher1994}. This allows to distinguish 
between 49s, 49p and 41d atoms. Electrons from 41d atoms arrive at the same 
time as electrons from 42p, so this method does not allow a distinction between 
these two states.  We use the final state fraction 
N$_{49p}/($N$_{49p} + $N$_{49s})$ as a measure of the dipole-dipole interaction.

It is important that both power supplies connected to the plates have a low 
noise level during the interaction time. This was realized by a metal shielding 
box around the home built pulse generator and an implemented line filter. The 
Agilent 33250A had a low noise level already. The total level of electric field 
noise is below \unitfrac[2]{mV}{cm}. We did not observe any influence of the 
rf-field on the SFI process as the amplitude of the rf-field is negligible 
compared with the magnitude of the SFI pulse, and as the frequency of the 
rf-field is too low to induce transitions to neighbouring Rydberg states.

\begin{figure}[ht]
	\centering
	\includegraphics[width=\columnwidth]{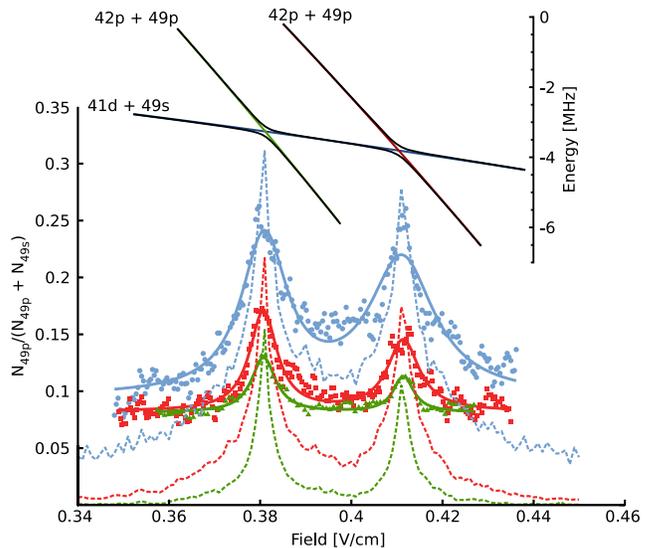}
	\caption{(Color online) Static field scans at 20 (blue circles) 30 (red 
	squares) and 40 (green triangles) $\mu$m separation between the 41d and the 
	49s volumes. Solid lines are fits with two independent Lorentz profiles. 
	Dashed lines are theoretical computations \cite{Ditzhuijzen2008}. The inset 
	shows the energy level diagram for two atoms at 25 $\mu$m. The energy scale 
	is relative to the energy of the 41d + 49s state at zero field.}
	\label{fig:fieldscan}
\end{figure}

We have greatly improved our setup by adding the ability to switch the coils 
generating the magnetic qua\-dru\-pole field for the MOT off during the 
dipole-dipole interaction time. Previously the magnetic field varied by 
\unit[$\approx0.5$]{G} over the length of the Rydberg volumes. This resulted in 
a broadening of the resonances by \unit[$\approx1.5$]{MHz}, due to Zeeman 
shifts of the involved states. For all results reported here the coils were 
switched off for \unit[6]{ms} in each \unit[100]{ms} cycle using a fast 
home-built IGBT switch. The current in the coils decays with a decay time of 
\unit[25]{$\mu$s}. Measurements are performed \unit[2.2]{ms} after the coils 
were switched. The magnetic field has been measured to be below 
\unit[0.14]{G} at this time. The remaining field is assumed to be due 
to background stray fields. The timing sequence of the experiment is shown in 
Fig. \ref{fig:timing}. 

Fig. \ref{fig:fieldscan} shows the 49p-fraction 
N$_{49p}/($N$_{49p} + $N$_{49s})$ as a function of static electric field for 
various separations of the dye laser beams creating the Rydberg atoms. The 
widths of the resonances have been improved significantly compared to the 
results presented in \cite{Ditzhuijzen2008} and are given in table 
\ref{tab:reswidth}. Resonance widths are extracted from the data by fitting two 
independent Lorentz profiles to the measurements. The improvement is clearly 
due to the switching of the magnetic field.  Our results are now in good 
agreement with the results of simulations \cite{Ditzhuijzen2008} 
which are also given in table \ref{tab:reswidth}. The remaining difference 
might be due to remaining variations in the electric field, or due to ions in 
the MOT-cloud. Note that a width of \unitfrac[5]{mV}{cm} corresponds to only 
\unit[661]{kHz} for resonance $F_\mathrm{a}$ and \unit[612]{kHz} for resonance 
$F_\mathrm{b}$. Both the simulations as well as the measurements yield widths 
bigger than the theoretical predictions of \unit[200]{kHz}. This discrepancy 
can be attributed to many-body effects as well as the finite width of the 
Rydberg volumes. Both effectes are taken into account in the simulations. 
In all further measurements discussed in this article a beam separation of 
\unit[25]{$\mu$m} is used.

\begin{table}
\centering
\begin{tabular}{|l|r@{.}l|r@{.}l|r@{.}l|r@{.}l|}
																		\hline
	& \multicolumn{4}{c|}{FWHM($F_\mathrm{a}$) \unitfrac{mV}{cm}} %
	& \multicolumn{4}{c|}{FWHM($F_\mathrm{b}$) \unitfrac{mV}{cm}} 	\\ \hline 
	& \multicolumn{2}{c|}{experiment} & \multicolumn{2}{c|}{theory} %
	& \multicolumn{2}{c|}{experiment} 	& \multicolumn{2}{c|}{theory} \\ \hline
\unit[20]{$\mu$m}	& 12&3(1.3)	& 11&0(1.3)	& 17&3(1.9)	& 11&5(1.5)	\\
\unit[30]{$\mu$m}	&  7&3(0.8)	& 8&1(0.9)	& 8&5(1.3)		& 7&8(1.0)\\
\unit[40]{$\mu$m}	&  6&4(0.9)	& 4&2(0.3)	& 5&3(1.2)		& 4&4(0.5)\\ \hline
\end{tabular}
\caption{Width (FWHM) of the resonances $F_\mathrm{a}$ and $F_\mathrm{b}$ for 
varying beam separations. Experimental and theoretical values are given. 
Theoretical data is taken from \cite{Ditzhuijzen2008}}
\label{tab:reswidth}
\end{table}

\section{Diabatic Switching of the Dipole-Dipole Interaction}
\begin{figure}[ht]
	\centering
	\includegraphics[width=\columnwidth]{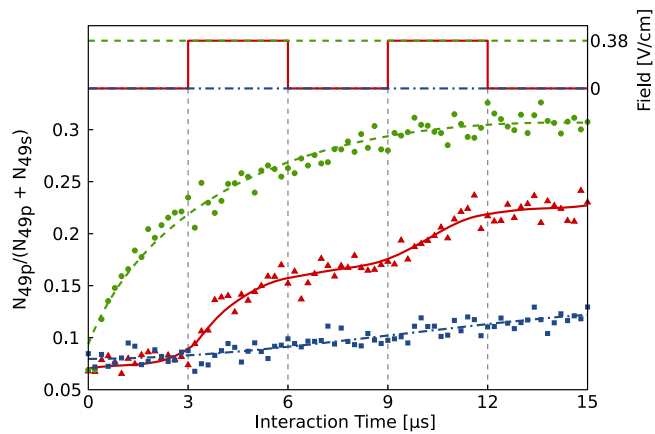}
	\caption{(Color online) 49p-fraction as a function of interaction time for 
	different fields $F = F_\mathrm{a}$ (green circles), $F = 0$ (blue squares) 
	and $F$ alternating between between $F_\mathrm{a}$ and 0 in the manner 
	depicted on top of the figure (red triangles). Lines depict a smooth spline 
	fitted to the data.}
	\label{fig:diab}
\end{figure}

In a first experiment we investigated the switching of the dipole-dipole 
interaction by repeatedly changing the electric field diabatically from the 
resonance $F_\mathrm{a}$ to a value 
\unitfrac[$F_{\mbox{\footnotesize{off}}} = 0$]{V}{cm}. The results are shown in 
Fig. \ref{fig:diab}. This figure depicts the 49p-fraction as a function of 
interaction time. The main curve in Fig. \ref{fig:diab} (red solid lines, 
triangles) shows the 49p-fraction for an electric field that is switched 
several times from \unitfrac[0]{V}{cm} to $F_\mathrm{a}$ and back again. 
The switching is done at a very fast rate of \unitfrac[76]{V}{cm$\mu$s}. This 
is much faster than the frequency of the quantum beat oscillations times the 
width of the resonance (\unitfrac[$2\times10^{-4}$~]{V}{cm$\mu$s}) ensuring 
that the system follows the field diabatically. Clearly the 49p-fraction rises 
during the times where $F = F_\mathrm{a}$ and stagnates when 
$F = F_{\mbox{\footnotesize{off}}}$. This shows how the dipole-dipole 
interaction can be switched on and off by rapidly changing the electric field. 
Reference measurements are performed for $F = F_\mathrm{a}$ at all times 
(green dashed curves, circles) and for $F=0$ at all times (blue dashed-dotted 
curves, squares). In the first case the 49p-fraction rises in the manner 
already measured in \cite{Ditzhuijzen2008}. When the field is off-resonant at 
all times there is a small background contribution due to black body radiation 
(see figure \ref{fig:energydensity}).

\section{Observation of rf-assisted multi-photon dipole-dipole transitions}

\begin{figure}[ht]
	\centering
	\includegraphics[width=\columnwidth]{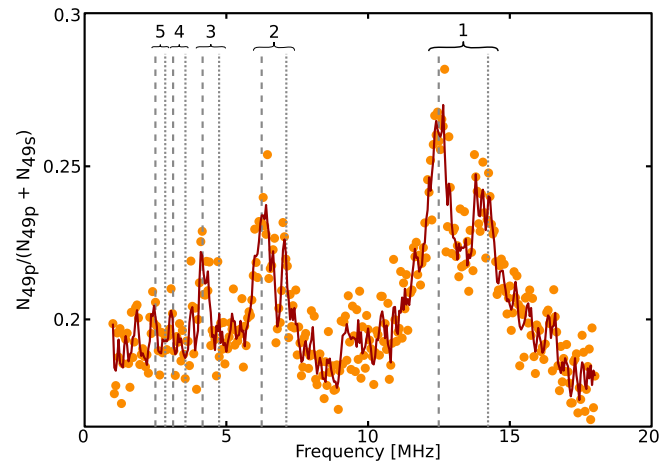}
	\caption{(Color online) Observed frequency dependence of rf-assisted 
	dipole-dipole transitions at $F_S$=\unitfrac[0.26]{V}{cm}. multi-photon 
	peaks with n=1--5 involved photons are clearly distinguishable, the number 
	of involved photons is denoted in the figure. The double peak structure 
	corresponds to the two resonances $F_\mathrm{a}$ ans $F_\mathrm{b}$.  
	The solid line is a smooth spline fitted to the data.}
	\label{fig:rfscan}
\end{figure}
Fig. \ref{fig:rfscan} shows the 49p-fraction as a function of the frequency 
of an external sinusoidal rf-field with constant amplitude and offset, here 
\unitfrac[$F_S = 260$]{mV}{cm} and \unitfrac[$F_{\mathrm{rf}} = 80$]{mV}{cm}. 
In this case amplitude and offset are chosen such that the field never reaches 
the resonance fields $F_\mathrm{a}$ or $F_\mathrm{b}$. The figure shows a 
series of double peaks; each sub-peak corresponds to one of the two resonances 
from the different $|m_j|$ substates, and each set corresponds to a different 
number of photons (N) needed to make the transition resonant, as noted in the 
figure. Up to 5-photon transitions are clearly visible in the figure. The 
frequency of each N-photon transition is 1/N-th of the one-photon transition 
frequency. The position of the peaks agrees very well with theoretical 
calculations based on eqn. \eqref{eqn:res} indicated by vertical lines 
(dashed: $F_\mathrm{a}$, dotted: $F_\mathrm{b}$) calculated for an effective 
field of \unitfrac[$F_{\mathrm{eff}} = 266.1$]{mV}{cm};

\begin{figure}[ht]
	\centering
	\includegraphics[width=\columnwidth]{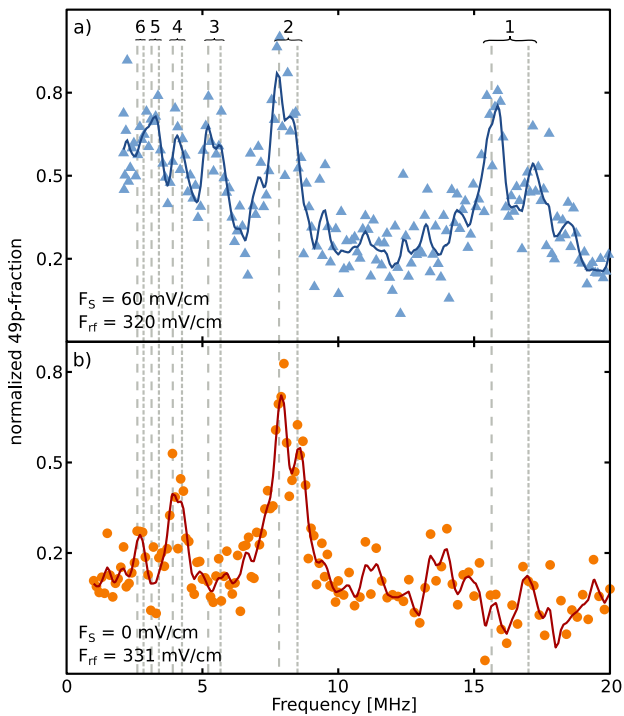}
	\caption{(Color online) Measurements at small static field $F_S$ (upper 
	panel) and at $F_S = 0$ (lower panel) showing selection rules prohibiting 
	odd photon number transitions (see text). The solid lines are a smooth 
	spline fitted to the data.}
	\label{fig:2rfscan}
\end{figure}

Fig. \ref{fig:2rfscan} again shows the 49p-fraction (normalized to the 
highest measured signal) as a function of frequency of the rf-field. The lower 
panel shows a frequency scan for \unitfrac[$F_S = 0$]{V}{cm} and 
\unitfrac[$F_{\mathrm{rf}} = 331$]{mV}{cm}, hence no static field is present in 
this measurement. The upper panel uses a small static field of 
\unitfrac[$F_S = 60$]{mV}{cm} and \unitfrac[$F_{\mathrm{rf}} = 320$]{mV}{cm} so 
both measurements have the same effective field 
\unitfrac[$F_{\mathrm{eff}} = 234$]{mV}{cm}. All peaks with odd photon number 
are clearly suppressed in the lower panel, but easily distinguishable in the 
upper panel. This can be understood on the basis of selection rules: The 
dipole-dipole interaction Eq. \eqref{eq:reaction2} requires that one unit 
of angular momentum is transferred from one atom to the other, 
$49\mathrm{s} \rightarrow 49\mathrm{p}$ requires $\Delta\ell = +1$ and 
$41\mathrm{d} \rightarrow 42\mathrm{p}$ requires $\Delta\ell = -1$. To achieve 
this, an odd number of photons must be involved in the process. Since there is 
already one photon transmitting the dipole-dipole interaction the number of 
external rf-photons must be even if the static field is zero. Already at static 
fields of only \unitfrac[60]{mV}{cm} this selection rule becomes invalid as a 
static-field photon can account for the missing angular momentum without 
transferring any energy. 

In Fig. \ref{fig:2rfscan} the importance of the ac Stark shift can clearly be 
observed. E.g.\ for resonance a, the energy difference between the sd- and the 
pp-state is \unit[24.53]{MHz} at \unitfrac[60]{mV}{cm}. Without ac Stark shift 
the 1,2 and 3-photon resonances are expected at 24.53, 12.26 and 
\unit[8.17]{MHz}. Including the ac Stark shift
\begin{equation}
	\Delta\omega = \frac{1}{4}\alpha F_{\mathrm{rf}}^2
\end{equation}
we get significantly different numbers, i.e. 15.64, 7.82 and \unit[5.31]{MHz}. 
These resonant positions clearly correspond to the observed peaks. The 
theoretical predictions including this ac stark shift are depicted in the 
figure as dash-dotted lines (resonance a) and dotted lines (resonance b).

\begin{figure}[ht]
	\centering
	\includegraphics[width=\columnwidth]{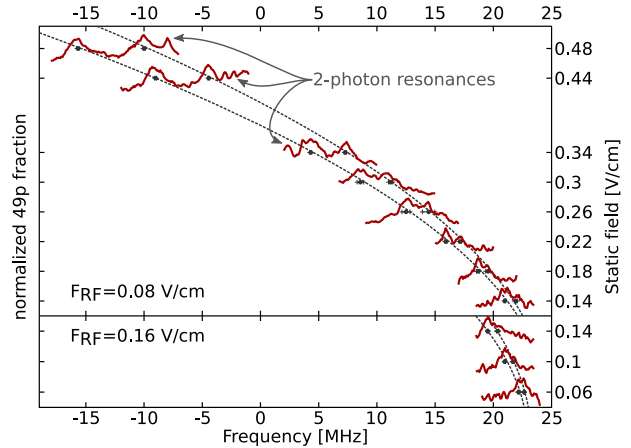}
	\caption{(Color online) $+1$ and $-1$ photon resonances at various effective 
	field values. In the lower panel (small static fields) the rf-amplitude had 
	to be increased leading to a larger ac-Stark shift. The fitted resonance 
	positions are marked by black dots. Some 2-photon resonances are also 
	visible and marked in the figure.}
	\label{fig:spectroscopy}
\end{figure}

In Fig. \ref{fig:spectroscopy} the 1-photon resonance is used to perform 
spectroscopy on the 2-atom states, measuring the energy difference between 
initial and final states for several values of static field offset. The figure 
shows the 49p-fraction as a function of both rf- and static field. The position 
of the peaks corresponds to the energy levels in Fig. \ref{fig:energy} as 
depicted by the dashed lines. These dashed lines are the same as the difference 
between the energy levels depicted in Fig. \ref{fig:energy}, but now with the 
x- and y axis interchanged. In the upper panel the amplitude of the 
radio-frequency field is \unitfrac[$F_{\mathrm{rf}} = 80$]{mV}{cm}. In the 
lower panel \unitfrac[$F_{\mathrm{rf}} = 160$]{mV}{cm} was used since the 
signal dropped too much for the lower amplitude at these small static field 
values. This leads to an appreciable contribution of the ac-Stark shift in the 
lower panel. The dashed lines show the theoretical resonance position taking 
ac-Stark shift into account on basis of Eq. \ref{eqn:res}. The $+1$ and $-1$ 
photon resonance position as a function of static field allows us to determine 
the total difference polarizability and difference energy between initial and 
final state. For the states investigated in this paper this yields 
\unit[$W_0 = 25.07 (10)$]{MHz}, 
\unitfrac[$\alpha_\mathrm{a}=347.2 (2.1)$]{MHz}{(V/cm)$^2$}  and 
\unitfrac[$\alpha_\mathrm{b}=300.5(1.9)$]{MHz}{(V/cm)$^2$}. Note that for these 
results the field strength was calibrated using static field scans such as 
shown in Fig. \ref{fig:fieldscan}. For accurate spectroscopy independent 
measurements of the electric field strength would be needed.

\section{Conclusion}
The results presented in this paper have been improved significantly compared 
to \cite{Ditzhuijzen2008} by switching off the magnetic field coils during 
measurements. The widths of the resonance peaks are now in good agreement with 
the simulations presented in \cite{Ditzhuijzen2008}. This improvement makes 
sub-MHz spectroscopy possible. We have shown that the dipole-dipole interaction 
can be manipulated by rapidly switching the electric field. Furthermore we have 
shown multi-photon resonances in rf-assisted dipole-dipole interaction with up 
to 5 involved photons where the interacting dipoles are localized in separate 
volumes. For large amplitudes of the rf-field significant ac-Stark shifts can 
be observed due to the quadratic field dependence of the involved states. The 
positions of the resonance peaks agree well with the theory presented here. The 
methods explored in this paper allow to perform spectroscopy on Stark states of 
two Rydberg atoms and we have determined the energy difference and difference 
in polarizability with good precision.

In this paper we have shown several versatile methods of manipulating the 
interactions between Rydberg atoms in separated volumes by external fields. 
These methods should prove to be valuable for the manipulation of dipole-dipole 
interaction between neutral atoms.

\begin{acknowledgments}
We thank Richard Newell for contributions to the experiment and Francis 
Robicheaux for stimulating discussions. This work is part of the research 
programme of the 'Stichting voor Fundamenteel Onderzoek der Materie (FOM)', 
which is financially supported by the 'Nederlandse Organisatie voor 
Wetenschappelijk Onderzoek (NWO)'.
\end{acknowledgments}

\appendix

\end{document}